\documentclass[doublecol]{epl2} 

\usepackage{times}
\usepackage[latin1]{inputenc}
\usepackage{graphicx}  
\usepackage{amsmath}
\usepackage{bm}
\usepackage{color}

\title{Ionic Cloud Distribution close to a Charged Surface 
       in the Presence of Salt}

\shorttitle{Strong coupling theory with added salt} 

\author{Olli Punkkinen\inst{1,2} \and Ali Naji\inst{3} \and Rudolf Podgornik\inst{4,5} 
\and Ilpo Vattulainen\inst{1,2,6} \and Per-Lyngs Hansen\inst{2,4} }

\shortauthor{O. Punkkinen \etal}

\institute{ 
  \inst{1} Laboratory of Physics and 
Helsinki Institute of Physics, Helsinki University of Technology - 
P.\,O. Box 1100, FI--02015 HUT, Finland\\
  \inst{2} Memphys\,--\,Center for Biomembrane Physics, 
Dept. of Physics and Chemistry, University of Southern Denmark - 
Campusvej 55, DK--5230 Odense M, Denmark\\
  \inst{3} Dept. of Physics, and Dept. of Chemistry and Biochemistry, University of California - 
Santa Barbara, CA 93106, USA \\
  \inst{4} Laboratory of Physical and Structural Biology, 
National Institutes of Health - Bldg. 9, MD 20892-0924, USA\\
  \inst{5} Dept. of Physics, Faculty of Mathematics and 
Physics, University of Ljubljana, and Dept. of Theoretical Physics, 
J. Stefan Institute - SI--1000 Slovenia\\
  \inst{6} Institute of Physics, 
Tampere University of Technology - P.\,O. Box 692, FI--33101 Tampere, Finland
}


\abstract{
Despite its importance, the understanding of ionic cloud distribution 
close to a charged macroion under physiological salt conditions 
has remained very limited especially for strongly coupled systems with, for instance, multivalent counterions. 
Here we present a formalism that predicts both counterion and coion distributions in the vicinity of a 
charged macroion for an arbitrary amount of  added salt and in both limits of mean field and strong coupling. 
The distribution functions are calculated explicitly for ions next to an infinite planar charged wall.
We present a schematic phase diagram identifying different physical regimes in terms of electrostatic 
coupling parameter and bulk salt concentration.
}

\begin{document}

\maketitle

\section{Introduction}

Electrostatic interactions play a key role in controlling  
solubility, structure and phase behavior of macroions in aqueous solutions 
\cite{charges:review,andelman:PB,israelachvili:intermolecular}. Examples of biologically 
relevant macroion systems are charged lipid bilayers such as those found in mitochondrial membranes 
that contain considerable amounts 
of anionic cardiolipins or plasma membranes rich in anionic phospholipids ({\em e.g.}  phosphatidylserines), 
stiff ({\em e.g.} DNA) or flexible ({\em e.g.} RNA) polyelectrolytes containing dissociated 
negatively charged phosphate groups, or charged polypeptides with a net charge depending on the dissociation 
equilibrium of various peptide moieties, as well as their complexes as encountered in the context of 
gene therapy \cite{nancy} or self-assembly of viruses \cite{bruinsma-virus}.  Instead of residing exactly 
on the charged surface, in order to minimize their electrostatic interaction energy, the counterions 
needed to neutralize these systems are distributed  some distance away as a consequence of their translational 
entropy as well as the screening effects due to residual co- and counterions. 
It is the nature of the spatial distributions of all these various mobile ion species 
in the vicinity of a charged macroion that presents the biggest challenge in understanding  
charged (bio)colloidal systems \cite{charges:review}. 
This is the problem that we scrutinize in what follows.

The traditional approach to charged macromolecular systems under 
{\em salt-free conditions} has been 
the Poisson--Boltzmann (PB) formalism, in which the Coulombic 
interaction between counterions is handled on a mean-field level \cite{andelman:PB,israelachvili:intermolecular}. 
However, in many biologically relevant  situations the PB approximation 
breaks down; examples include most prominently multivalent counterions and highly charged surfaces. 
The most dramatic indication of this breakdown is the existence of attractive interactions 
between like-charged macroions 
\cite{kjellander:integral,wennerstrom:water-uptake,kekicheff:attraction-apparatus, lowen:DNA-attraction}  
which on the PB level are known to be repulsive \cite{neu}. 
Consequently, there have been a number 
of attempts to assess corrections to the PB theory using, 
{\em e.g.}, correlated density fluctuations around the mean-field 
distribution or additional non-electrostatic interactions 
\cite{andelman:non-ele,kjellander:integral,orland:beyond-PB,podgornik,pincsu:fluctuation}. 
An alternative approach has been pioneered by Rouzina and 
Bloomfield \cite{rouzina:sc} and elaborated by Shklovskii {\em et al.} \cite{shklovskii}, 
and later by Netz {\em et al.} \cite{charges:review,netz:sc,netz:simulation,naji-review}. 
This approach leads to a new description of a system composed of a charged macroion and mobile counterions called 
the strong coupling (SC) theory. This description was shown to become exact in the limit of high surface 
charge, multivalent counterions, or low temperature \cite{netz:sc}, clearly opposite to the PB limit, 
asymptotically valid in the limit of low surface charge, monovalent counterions, or high temperatures. 
The PB and the SC theories thus asymptotically embrace 
all possible scenarios in the no-salt case. The SC 
theory has been applied with notable success to the 
case of charged macroscopic  surfaces of various geometries with counterions \cite{naji-review}.

However, under physiologically relevant conditions, the situation is 
considerably more complicated. Biological systems always 
contain significant concentrations of excess salts, which affect 
or quite often even govern their behavior. Overall, in biological 
conditions the reservoir salt concentration is typically of the order 
of $[\mathrm{Na}^{+}] = 100$~mM. It is thus obvious that the 
average separation between nearest salt ions is small, implying that
there is no justification to disregard the effects of salt, 
as opposed to the effects of counterions, in a consistent statistical-mechanical treatment of such 
systems. The traditional approach in this context has been 
the Debye--H\"uckel (DH) theory, related to the PB formalism, 
that applies in the limit of small overall charges in the system 
\cite{israelachvili:intermolecular,debye:screening,podgornik}.  Apart from this limiting case, 
a coherent theoretical description including the SC limit has been missing.

In this work we propose a consistent and systematic approach to charged systems, 
composed of fixed macroions, multivalent counterions (of charge valency $+q_c$) 
as well as added salt (of cationic and anionic charge species of valency $+q_+$ and $-q_-$)
in chemical equilibrium with a bulk reservoir. 
We consider the influence of added salt on ion distributions on the PB level as well 
as in the SC limit. The results are shown to fill the gap between the no-salt and high salt limits. 
In particular, we show how certain divergencies may be removed and 
normalizable ion density profiles may be obtained in the SC limit. Explicit calculations are carried out 
for ions at an infinite planar charged wall, where we also present subleading corrections to the SC results.

Before we formulate the partition function for an interacting 
system of this type in a general form, let us introduce the relevant 
length scales and parameters of the problem, and show by scaling arguments 
what one should expect from a more rigorous theory. First, we focus on counterions. An 
important length scale in the problem  is the \emph{Gouy--Chapman} (GC) length 
$\mu = 1/(2\pi q_c\sigma_s l_\mathrm{B})$. It measures the typical distance from 
a charged macroion surface (of surface charge density $-\sigma_s$) at which the electrostatic 
potential energy of a counterion interacting with the surface matches the thermal energy 
$k_{\mathrm{B}}T$. Here, $l_\mathrm{B} = e^2/(4\pi\epsilon k_\mathrm{B} T)$ 
is the \emph{Bjerrum length}, the distance at which the 
interaction between two unit charges equals thermal energy; in water 
$l_\mathrm{B} \simeq 0.7$~nm. The ratio between these two length scales
 yields an important dimensionless parameter, namely,  
the {\em electrostatic coupling parameter} \cite{netz:sc}
\begin{equation}
\label{eq:Xi}
\Xi \equiv q_c^2\frac{l_\mathrm{B}}{\mu} = 2\pi q_c^3 l_\mathrm{B}^2 \sigma_s, 
\end{equation}
embodying the relative contribution of  ion-ion {\em vs.} ion-surface interactions. 
A related length scale is the lateral distance between counterions 
forming a strongly correlated quasi-2D layer close to a charged 
surface in the SC regime \cite{charges:review,naji-review}, {\em i.e.}
$a_{\perp} = \sqrt{(q_c/\pi\sigma_s)} = \sqrt{2\Xi}\mu$. 
This lateral distance is important, because in the SC 
regime we can express the concentration of the counterion 
layer as $n_c = {\sigma_s}/({q_c\mu}) = 
{1}/({\pi a_{\perp}^2\mu})$  \cite{rouzina:sc}. 
The counterion concentration within this 
strongly correlated diffuse layer should be independent 
of the bulk ion concentration, as long as the ionic strength, 
defined as $I = \frac{1}{2}\sum_{i=\pm} \Lambda_i q_i^2$, is significantly 
smaller than $n_c$, {\em i.e.} $I \ll n_c$. Here $ \Lambda_i$ and $q_i$ 
are the fugacities (to be defined later) and valencies of reservoir salt ions. 
This means that the counterion concentration on a surface 
should be much larger than the reservoir salt concentration.

In a system that contains added salt there exists of 
course another independent length scale, the
\emph{Debye screening length} $l_\mathrm{DH}^2 = 1/(8 \pi I l_\mathrm{B})$ 
being related to the ionic strength $I$.  
It measures the distance at which the Coulombic interaction 
between two point charges is screened out \cite{debye:screening}. The effects of salt 
thus become unimportant when $\kappa\mu \ll 1 $, where $\kappa = l_\mathrm{DH}^{-1}$ 
denotes the inverse Debye screening length. 
Later on we will see that it is really the coupling 
constants $\kappa\mu$ and $\Xi$ that uniquely determine 
the phase diagram of the system.

\section{Model}

In what follows (details will be given elsewhere \cite{olli:sc}), 
we present a general formalism for a negatively charged fixed macroion 
interacting with $N_c$ counterions, $N_{+}$ positive salt ions, 
and $N_{-}$ negative salt ions in equilibrium with a bulk reservoir. 
The $N_c$ counterions are assumed 
to have a valency which may be greater than that of salt ions and 
therefore have to be treated differently. 
To proceed formally, we introduce the canonical partition function for 
the mobile charged species, confined to some arbitrary region 
around a fixed charged macroion characterized by charge density 
$\sigma(\mathbf{r})$, {\em i.e.} 
\begin{equation}
\label{eq:cano_part1}
Z = \bigg[\prod_{\alpha}\frac{1}{N_{\alpha}!} \prod_{j_{\alpha}=1}^{N_{\alpha}}
  \int\! \mathrm{d}\mathbf{r}_{j_{\alpha}} 
  \Omega(\mathbf{r}_{j_{\alpha}}) \bigg]
\,e^{-\mathcal{H}}. 
\end{equation}
Here, $\Omega(\mathbf{r})$ restricts the positions of mobile ions 
to the region of space available to them.  The 
index $\alpha=\{c, \pm\}$ stands for counterions ($c$), positive ($+$) as well as 
negative ($-$) salt ions. In what follows, we assume that the macroion charge density $\sigma(\mathbf{r})$
is negative and confined to the macroion surface with surface charge density $\sigma_s$.
Introducing the density operator for each 
ion type $\hat{\rho}_{\alpha}(\mathbf{r}) = 
\sum_{j=1}^{N_{\alpha}}\delta(\mathbf{r}-\mathbf{r}_{j})$, 
the Hamiltonian can be written in units of $k_\mathrm{B} T$ as 
\begin{eqnarray}
\label{eq:hamil1}
\mathcal{H}\!\!\!\! & =& \!\!\!\!
	\frac{1}{2}\int\!\!\int\mathrm{d}\mathbf{r}\,\mathrm{d}\mathbf{r'}
		\big[q_c\hat{\rho}_c(\mathbf{r}) + q_{+}\hat{\rho}_{+}(\mathbf{r}) 
			- q_{-}\hat{\rho}_{-}(\mathbf{r}) -\sigma(\mathbf{r})\big] \nonumber\\ 
	& & \!\!\!\!\times\, v(\mathbf{r} - \mathbf{r'})
		\big[q_c\hat{\rho}_c(\mathbf{r'}) + q_{+}\hat{\rho}_{+}(\mathbf{r'}) - 
			q_{-}\hat{\rho}_{-}(\mathbf{r'}) -\sigma(\mathbf{r'})\big] \nonumber\\ 
	& & \!\!\!\! + \sum_{\alpha}\big[-\frac{l_\mathrm{B}}{2}N_{\alpha}q_{\alpha}^2 v(\mathbf{0}) 
		- \int\mathrm{d}\mathbf{r}\,h_{\alpha}(\mathbf{r})\hat{\rho}_{\alpha} 
			(\mathbf{r})\big], 
\end{eqnarray}
where $v(\mathbf{r}) = l_{\mathrm{B}}/|{\mathbf r}|$ is the Coulomb interaction, and the 
generating fields $h_{\alpha}(\mathbf{r})$ have been added to 
calculate ion distributions by taking functional derivatives. 
Here we have also explicitly subtracted the infinite self-energies.

At this stage, we proceed by applying the Hubbard--Stratonovich 
transformation \cite{hubbard:stratonovich}, the purpose of which is to get rid of the quadratic 
density terms in $\mathcal{H}$ at the expense of introducing the fluctuating electrostatic potential 
field, $\phi(\mathbf{r})$. 
This is followed by a Legendre transform to grand-canonical ensemble, 
where the number of ions $\{N_c, N_\pm\}$is replaced by their fugacities 
$\{\tilde \Lambda_c, \tilde \Lambda_\pm\}$. Next follows the 
crucial step which makes the field-theoretic partition function 
convergent. We add the exponential of the following expression \cite{netz:virial,olli:sc}
\begin{equation} 
\int\mathrm{d}\mathbf{r}\,\Omega(\mathbf{r})\left(
	\sum_{\alpha}\tilde{\Lambda}_{\alpha} 
		+ \frac{1}{2}\sum_{i=\pm}\tilde{\Lambda}_i
			[\phi(\mathbf{r})^2 + v_\mathrm{DH}(\mathbf{0})]\right),  
\label{renorm}
\end{equation}
to the partition function and subtract it perturbatively. 
Here $ v_\mathrm{DH}(\mathbf{r})$ is the inverse of the DH operator 
$4\pi l_{\mathrm{B}}~v_\mathrm{DH}^{-1}({\mathbf{r}} - {\mathbf{r}}') = 
(-\nabla^2 + \kappa^2) ~\delta({\mathbf{r}} - {\mathbf{r}}')$, 
and of course corresponds to the screened DH interaction potential.  
Expressing the partition function in this form accomplishes three tasks: first, it 
removes the bulk density values of all ion types in order to make 
the one-particle densities finite. Second, the screening factor $\kappa^2\phi^2$ 
makes the range of interaction between all the charges finite; and finally, the infinite self-energies 
cancel another set of divergencies present in the partition function.
Rescaling all lengths by the GC length 
$\mu = 1/(2\pi l_\mathrm{B} q_c \sigma_s)$ according to 
$\mathbf{r} = \mu\tilde {\mathbf r}$, one ends up with an exact field-theoretic representation for
the grand-canonical partition function 
${\mathcal Q}=\int{\mathcal D}\phi\,\exp(-\tilde {\mathcal H}/\Xi)$, 
where $\tilde {\mathcal H}$ is the rescaled effective Hamiltonian \cite{olli:sc} 
\begin{equation}
\label{eq:hamil_scal}
\begin{split}
\tilde{{\mathcal H}}[\phi] &= 
	{\textstyle\frac{1}{2}} q_c^2\int\!\mathrm{d}\tilde{\mathbf{r}}\,\mathrm{d}\tilde{\mathbf{r}}'
		\phi(\tilde{\mathbf r})\tilde v_\mathrm{DH}^{-1}(\tilde{\mathbf{r}} - 
			\tilde{\mathbf{r}}') \phi(\tilde{\mathbf{r}}') \\ 
	& - i\frac{q_c}{2\pi}\int\!\mathrm{d}\tilde{\mathbf{r}} \,
		\phi(\tilde{\mathbf{r}})\tilde \sigma(\tilde{\mathbf{r}}) \\ 
	& - \frac{\Lambda_{c}}{2\pi}\int\!\mathrm{d} 
		\tilde{\mathbf{r}}\, \tilde{\Omega}(\tilde{\mathbf{r}})
			[e^{h_c(\tilde{\mathbf{r}}) - iq_c\phi(\tilde{\mathbf{r}}) + 
				\Xi v_\mathrm{DH}(\tilde{\mathbf{0}})/2} - 1] \\ 
	& - \frac{\Lambda_{+}} 
		{2\pi}\int\!\mathrm{d}\tilde{\mathbf{r}}\,
		\tilde{\Omega}(\tilde{\mathbf{r}})\,Q_{+}(\tilde{\mathbf{r}}) 
		- \frac{\Lambda_{-}}{2\pi}\int\!\mathrm{d} 
		\tilde{\mathbf{r}}\,\tilde{\Omega}(\tilde{\mathbf{r}})\,Q_{-}(\tilde{\mathbf{r}}).
\end{split}
\end{equation}
Here we introduce a shorthand $Q_{i}(\tilde{\mathbf r}) = 
\exp[h_{i}(\tilde{\mathbf r}) - iq_{i}\phi(\tilde{\mathbf r}) 
+ (\Xi/2)(q_{i}^2/q_c^2) \tilde v_{\mathrm{DH}}(\tilde{\mathbf 0})/2] - 
\exp[({\Xi}/{2})({q_{i}^2}/{q_c^2})\Delta v_0] 
+ ({q_{i}^2}/{2})\phi(\tilde{\mathbf r})^2 - 
({\Xi}/{2})({q_{i}^2}/{q_c^2})\tilde  v_\mathrm{DH}(\tilde{\mathbf 0})$, 
while also rescaling the fugacities such that 
$\Lambda_{\alpha} = (2\pi \Xi \mu^3)\tilde{\Lambda}_{\alpha}$. 
Here we also defined $4\pi[-\tilde{\nabla}^2+(\kappa\mu)^2]\tilde v_{\mathrm{DH}}^{-1}(\tilde{\mathbf{r}}) 
= \delta(\tilde {\mathbf r}- \tilde {\mathbf r}')$, and 
$\Delta v_{0} = \tilde v_c(\tilde{\mathbf 0}) - \tilde v_\mathrm{DH}(\tilde{\mathbf 0})$. 
The expectation values of different ion densities can be 
calculated by taking a functional derivative of the 
grand-canonical free energy with respect to 
the generating field $h_{\alpha}(\tilde{\mathbf r})$, 
$\langle{\rho_{\alpha}(\tilde{\mathbf r})\rangle} = 
\delta \ln{\mathcal{Q}}/\delta h_{\alpha}(\tilde{\mathbf r})\mu^3|_{h_{\alpha}=0}$, 
giving rise to the rescaled densities 
\begin{equation}
\label{eq:density}
\langle\tilde{\rho}_{\alpha}(\tilde{\mathbf r})\rangle = 
	\frac{\langle\rho_{\alpha}(\tilde{\mathbf r})\rangle}{2\pi l_\mathrm{B} \sigma_s^2} = 
		\Lambda_{\alpha}\tilde{\Omega}(\tilde{\mathbf{r}})\langle e^{-iq_{\alpha}\phi(\tilde{\mathbf{r}})}\rangle,
\end{equation}
where we have redefined $\Lambda_{\alpha}\rightarrow \Lambda_{\alpha}e^{\frac{\Xi}{2}\frac{q_{\alpha}^2}{q_c^2}v_\mathrm{DH}(\tilde{\mathbf{0}})}$. 
The normalization condition for the ion distributions then follows as 
$\int\mathrm{d}\tilde{\mathbf{r}}\,
[q_c\tilde\rho_c(\tilde{\mathbf{r}}) + q_+\tilde\rho_{+}(\tilde{\mathbf{r}}) - 
q_-\tilde\rho_{-}(\tilde{\mathbf{r}})] = q_c\int\mathrm{d}\tilde{\mathbf r}\,
\tilde\sigma(\tilde{\mathbf{r}})$, where $\tilde\sigma(\tilde{\mathbf{r}}) = (\mu/\sigma_s)\sigma( {\mathbf{r}})$.  
This corresponds to the overall electroneutrality of the system.

\section{Results}

\begin{figure}[tb] 
\begin{center} 
\includegraphics[width=7cm]{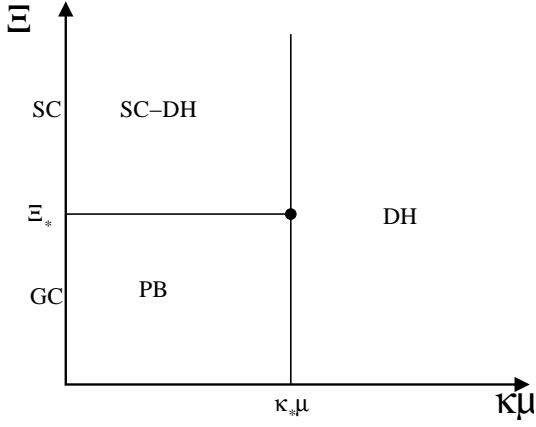} 
\end{center} 
\caption{\label{fig:phase_diagram}
Schematic phase-diagram representing different regimes of behavior
as a function of the two coupling parameters $\Xi$ and $\kappa\mu$. The crossover
from PB to SC-DH regime takes place at sufficiently small salt concentration and
by increasing the electrostatic coupling parameter beyond a typical value of  $\Xi_\ast \sim 10$ \cite{netz:simulation}.
While, for elevated salt concentration $\kappa\mu > \kappa_\ast\mu \sim 1$, DH is the
dominant regime.
}
\end{figure}

Employing the Hamiltonian in eq.~(\ref{eq:hamil_scal}), we 
next make a full classification of possible limiting cases 
in terms of the coupling parameters $\{\Xi, \kappa\mu\}$; 
see fig.~\ref{fig:phase_diagram}. 
These limiting cases are
\begin{itemize}
\item{(i)}~First, in the limit $\Xi \ll 1$, 
we find the familiar PB theory. This well-known regime 
\cite{stu:samuel,andelman:PB,israelachvili:intermolecular} is characterized 
by many-body interactions among uncorrelated ions. 
Mathematically, it follows from the saddle-point equation for $\tilde{{\mathcal H}}[\phi]$, yielding
the so-called Poisson--Boltzmann equation for the mean electrostatic 
potential $\phi_{\mathrm{PB}}$, {\em i.e.} 
\begin{equation}
q^2_c \tilde{\nabla}^2 \phi_{\mathrm{PB}} = 
		- 2 \sum_{\alpha} s_\alpha q_{\alpha}
		\Lambda_{\alpha}\tilde{\Omega}(\tilde{\mathbf{r}})\,e^{-s_\alpha
			 q_{\alpha} \phi_\mathrm{PB}},
\end{equation}
where $s_\alpha=\pm 1$ denotes the positive or negative sign of ions.  
Weakly correlated Gaussian fluctuations around the saddle-point solution may be captured by 
a loop-expansion in powers of $\Xi$ \cite{netz:sc}. This regime  separates into two sub-regimes 
according to the value of $\kappa\mu$, namely, the GC ($\kappa\mu \rightarrow 0$) 
and the DH ($\kappa\mu \rightarrow \infty$) regimes, which correspond to nonlinear and linear
PB equations, respectively. The free energy 
of the system, $F$, is in both cases found to be a decreasing function of bulk salt concentration, 
but an increasing function of $\Xi$, scaling as $F_\mathrm{PB} \sim \Xi$ \cite{olli:sc}. 
These regimes have been throughly studied before \cite{andelman:PB,israelachvili:intermolecular} 
and will not be considered here any further. 

\item{(ii)}~In the limit 
$\Xi \gg 1$, we end up with the SC theory, representing a highly correlated system. 
Mathematically, this follows from a virial expansion in powers of $\Xi^{-1}$ \cite{netz:sc}. 
Since the electrostatic interaction
between individual mobile ions and the charged macroion dominates 
over the ion-ion interactions, the leading-order SC theory comprises a single-particle description 
with an effective ion-surface interaction potential of the form
\begin{equation}
\tilde u(\tilde{\mathbf{r}}) = - \frac{1}{2\pi} \int d\tilde{\mathbf{r}}' \tilde\sigma(\tilde{\mathbf{r}}') \left[ \frac{e^{- \kappa\mu \vert \tilde{\mathbf{r}} - \tilde{\mathbf{r}}'\vert}}{\vert \tilde{\mathbf{r}} - \tilde{\mathbf{r}}'\vert} - \frac{e^{- \kappa\mu \vert \tilde{\mathbf{r}}'\vert}}{\vert \tilde{\mathbf{r}}' \vert}\right].
\label{eq:utilde}
\end{equation}
By adding enough bulk salt, the SC ionic distributions are destroyed, 
and the system ends up in the DH-regime, where electrostatic interactions 
are completely screened out. This shows up in 
the partition function as 
$\Lambda_{\pm} \sim (\kappa\mu)^2$ scaling \cite{olli:sc}, since the salt part of 
$\tilde {\mathcal H}[\phi]$ eventually starts to dominate. The 
free energy is again a decreasing function of salt concentration, 
but decreasing as $F_\mathrm{SC} \sim 1/\Xi$.
\end{itemize}

The crossover from the PB to SC regime for the no-salt case has been extensively 
studied in the simulations \cite{netz:simulation}, where the strong-coupling features are
 shown to set in at intermediate couplings about $\Xi_\ast \sim 10$. For the case 
with added salt such an analysis remains to be done. 

The asymmetric expansion of the partition function to the second 
order in $\Lambda_c/\Xi$ and to the first order in 
$\Lambda_{\pm}/\Xi \sim (\kappa\mu)^2/\Xi$, is 
equivalent to the virial expansion used in the SC limit without 
added salt, together with the Mayer--Friedman 
resummation of the grand-canonical partition function 
for the simple salt ions, giving 
rise to the screened Debye--H\" uckel potential 
\cite{mayer:virial,friedman}. Therefore we propose to call 
this expansion the {\em Strong Coupling with Debye--H\"uckel} (SC-DH) 
theory, see fig.~\ref{fig:phase_diagram}, identified already 
by Boroudjerdi {\em et al.} \cite{charges:review}. 
We also expand different fugacities in powers of $1/\Xi$, as 
$\Lambda_{\alpha} = \Lambda_{\alpha}^{0} + \Lambda_{\alpha}^{1}/\Xi 
+ \cdots $, which allows us to avoid divergencies arising from 
the second virial coefficient \cite{netz:sc,netz:simulation}.

Next we evaluate the ion densities to the lowest 
order in $\Lambda_{\alpha}/\Xi$, and obtain ion density expansions as
$\langle\tilde{\rho}_{\alpha}\rangle = \tilde{\rho}_{\alpha}^0 + 
\tilde{\rho}_{\alpha}^1/\Xi+\cdots$ in the SC limit. 
This density expansion, listed below for all the ionic species 
(\ref{eq:rho_c_legendre})-(\ref{eq:rho_-_legendre}), is in 
fact the main result of this paper. For counterion 
density  we get 
\begin{equation}
\label{eq:rho_c_legendre}
\begin{split}
\langle\tilde{\rho}_c(\tilde{\mathbf r})\rangle =& 
	\Lambda_c^{0} \,\tilde{\Omega}(\tilde{\mathbf{r}})\,e^{-\tilde{u}(\tilde{\mathbf{r}})} 
		+ \frac{1}{\Xi}\tilde{\Omega}(\tilde{\mathbf{r}})\,e^{-\tilde{u}(\tilde{\mathbf{r}})} \\ 
	& \left\{\Lambda_c^1 - \frac{(\Lambda_c^{0})^2}{2\pi}
				\int\!\mathrm{d}\tilde{\mathbf{r}}'\,\tilde{\Omega}(\tilde{\mathbf{r}}')\,
				[e^{-\tilde{u}(\tilde{\mathbf{r}}')}- e^{-\tilde{u}(\infty)}] \right. \\ 
	& \left. \times\left[1 - e^{-\Xi v_\mathrm{DH}(\tilde{\mathbf{r}} - 
		\tilde{\mathbf{r}}')}\right]\right\} 
			+ \mathcal{O}\left(\Xi^{-2},[\kappa\mu]^2/\Xi\right) 
\end{split}
\end{equation}
and for salt-ion  densities 
\begin{equation}
\label{eq:rho_+_legendre}
\begin{split}
  \langle\tilde{\rho}_+(\tilde{\mathbf r})\rangle =& 
  \Lambda_{+}^{0}\,\tilde{\Omega}(\tilde{\mathbf{r}})\,
  e^{-\frac{q_{+}}{q_c}\tilde{u}(\tilde{\mathbf{r}})} 
  + \mathcal{O}\left(\frac{[\kappa\mu]^2}{\Xi}\right)
\end{split}
\end{equation}
\begin{equation}
\label{eq:rho_-_legendre}
\begin{split}
  \langle\tilde{\rho}_-(\tilde{\mathbf r})\rangle =& 
  \Lambda_{-}^{0}\,\tilde{\Omega}(\tilde{\mathbf{r}})\,
  e^{+\frac{q_{-}}{q_c}\tilde{u}(\tilde{\mathbf{r}})} 
  + \mathcal{O}\left(\frac{[\kappa\mu]^2}{\Xi}\right).  
\end{split}
\end{equation}
The second order terms can also be included in 
the above two equations, but they do not 
provide any further relevant insight \cite{olli:sc}. 
Note that the key factor in the above expressions is the 
rescaled single-particle interaction term 
$\tilde{u}(\tilde{\mathbf{r}})$, which corresponds to 
a single ion interacting with the charged macroion, 
eq.~(\ref{eq:utilde}). No assumption has been made thus far about the geometry 
or symmetry properties of the macroion and the expressions 
(\ref{eq:rho_c_legendre})-(\ref{eq:rho_-_legendre}) have a completely general validity 
in the SC limit. These formulas illustrate explicitly 
that we have an expansion in terms of the single-particle 
density differences 
$[e^{-\tilde{u}(\tilde{\mathbf{r}})} - e^{-\tilde{u}(\infty)}] 
\sim  [\rho(\tilde{\mathbf{r}})-\rho(\infty)]$, 
stemming from the properly renormalized partition function
with the counter-terms eq.~(\ref{renorm}). These density 
differences are perfectly normalizable as demanded by the 
electroneutrality condition, whereas the densities 
themselves are not.

The SC-DH can be evaluated in closed form only in the case of a single infinite plate, 
{\em i.e.} in the plane-parallel geometry 
with $\tilde \sigma(\tilde {\mathbf r})=\delta(\tilde z)$ and counterions and salt ions
being present on both sides of the plate. From eq.~(\ref{eq:rho_c_legendre}) we obtain to the leading order 
the SC counterion density as a function of the perpendicular distance 
$\tilde{z}>0$ from the wall, {\em i.e.}
\begin{equation}
\label{eq:density_c0normal_1plate}
\tilde{\rho}_c^{0}(\tilde{z}) =
\Lambda_{c}^{0}\exp\big\{+[e^{-\kappa\mu \tilde{z}}-1]/\kappa\mu\big\}
\end{equation}
where the leading-order fugacity coefficient is given by 
\begin{equation}
\label{eq:fugacity_c0normal_1plate}
\begin{split}
  \Lambda_c^{0} =&
  \kappa\mu\Big[-2\{e^{(q_{+}+q_{-})/q_c\kappa\mu}q_{+}+q_{-}\} + \Big. \\ 
    &\Big.	q_c\kappa\mu e^{q_{-}/q_{c}\kappa\mu}\{\mathrm{Ei}[q_{+}/q_{c}\kappa\mu]-
    \mathrm{Ei}[1/\kappa\mu]+\log{[q_{-}/q_{+}]}\}\Big]/\\ 
  & \Big[2q_{-}\{\mathrm{Ei}[q_{+}/q_{c}\kappa\mu]-\mathrm{Ei}[1/\kappa\mu]+\log{[q_{c}/q_{+}]}\} \Big. \\
    &\Big. - 2q_{+}e^{(q_{+}+q_{-})/q_c\kappa\mu} \Big. \\ 
& \Big.\{-\mathrm{Ei}[-q_{-}/q_{c}\kappa\mu]+\mathrm{Ei}[1/\kappa\mu]+\log{[q_{-}/q_{c}]}\}\Big] \\
  & \xrightarrow{\kappa\mu \ll 1} 1-\kappa\mu-[1 + \frac{1}{2}\frac{q_c^2}{q_{+}^2}](\kappa\mu)^2 + \mathcal{O}([\kappa\mu]^3).
\end{split}
\end{equation}
Thus, in the limit $\kappa\mu \rightarrow 0$ counterion density approaches 
\begin{equation}
\label{eq:limit_density}
\begin{split}
\tilde{\rho}_c^{0}(\tilde{z}) \xrightarrow{\kappa\mu \ll 1}& e^{-\tilde{z}}
\left[1 - \kappa\mu(1 - \frac{1}{2}\tilde{z}^2) - (\kappa\mu)^2\times \right. \\
  & \left. [1 + \frac{1}{2}\frac{q_c^2}{q_{+}^2} - \frac{1}{2}\tilde{z}^2 
    + \frac{1}{6}\tilde{z}^3 - \frac{1}{8}\tilde{z}^4] + 
\mathcal{O}([\kappa\mu]^3)\right].
\end{split}
\end{equation}
This explicitly shows that we find the no-salt SC result 
in the limit $\kappa\mu \rightarrow 0$ \cite{netz:sc}, 
{\em i.e.} $\tilde{\rho}_c^{0}(\tilde{z})\rightarrow e^{-\tilde z}$. 
Note that the salt correction is negative for small distances, {\em i.e.}, 
it reduces the density close to the charged wall. This 
is expected intuitively since the  counterions tend 
to escape further away  from the wall due to 
the reduced interaction in the presence of Debye screening. 
In the limit $\kappa\mu\rightarrow \infty$, we find the usual DH expression, i.e.
$\tilde{\rho}_c^{0}(\tilde{z})\simeq \tilde{\rho}_c^b(1+e^{-\kappa\mu\tilde z}/\kappa\mu)$, where 
$\tilde{\rho}_c^b=\Lambda_{c}^{0}e^{-1/\kappa\mu}$ is the bulk density.

\begin{figure}[tb] 
\begin{center} 
\includegraphics[width=7.5cm]{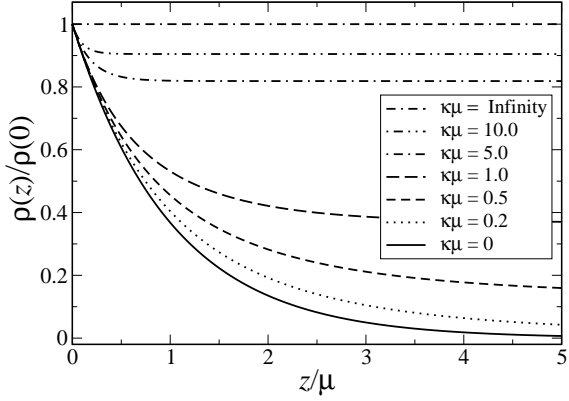} 
\end{center} 
\caption{\label{fig:density}
Leading order counterion density in the SC limit for different values of $\kappa\mu$. 
For the sake of representation, we show the ratio of density to its contact value, 
$\rho_{c}^{0}(z)/\rho_{c}^{0}(0)$, as obtained from eq. (\ref{eq:density_c0normal_1plate}). }
\end{figure}

Therefore, the excess density effectively crosses over from one exponentially decaying form,
i.e. the no-salt SC expression $e^{-z/\mu}$ with the decay length, $\mu$
(in actual units), to another exponentially decaying form, i.e. the DH expression with the decay length $\kappa^{-1}$.  
In fig.~\ref{fig:density} we show plots of the zeroth order ion-density 
$\tilde{\rho}_{c}^{0}(\tilde z)$ relative to the contact value $\tilde{\rho}_{c}^{0}(0)$ 
for different values of $\kappa\mu$, clearly attesting to the fact that our results nicely
interpolate between the well-known DH- and SC-regimes.

The first order correction in $1/\Xi$ for a planar charged surface 
follows again from  eq.~(\ref{eq:rho_c_legendre}).  The resulting expression can be 
given in terms of hypergeometric functions, but here we only 
present the result for $\kappa\mu \ll 1$, {\em i.e.} \cite{olli:sc} 
\begin{equation}
\label{eq:density_c1normal_1plate}
\tilde{\rho}_c^{1}(\tilde{z}) = e^{-\tilde{z}} \big[(\frac{1}{2}\tilde{z}^2 - \tilde{z}) - 
		\kappa\mu(2\tilde{z} + 1)\big] + \mathcal{O}([\kappa\mu]^2),   
\end{equation}
which again exhibits a smooth transition to the no-salt case \cite{netz:sc}, 
and shows that also the first-order correction to the density decreases close to the 
charged macroion surface when salt is introduced. 
Both, the zeroth- as well as the first-order counterion density profiles thus show a smooth transition 
to the  no-salt case attesting to the consistency of our formulation.

The concentrations of the salt ions can be obtained in the same way from 
eqs.~(\ref{eq:rho_+_legendre}) and (\ref{eq:rho_-_legendre}) as 
\begin{eqnarray}
\label{eq:bulk_salt_1plate}
\tilde{\rho}_{+}^{0}(\tilde{z}) & = &  \Lambda_{+}^{0} \,
	\exp\big\{+[e^{-\kappa\mu \tilde{z}}-1]q_{+}/(q_{c}\kappa\mu)\big\}, \\ 
\tilde{\rho}_{-}^{0}(\tilde{z}) &=& 
 	\Lambda_{-}^{0}\,\exp\big\{-[e^{-\kappa\mu \tilde{z}}-1]q_{-}/(q_{c}\kappa\mu)\big\}, 
\end{eqnarray}
for $\tilde{z}>0$, which obey the overall electroneutrality condition infinitely far 
away from the wall in the form 
\begin{equation}
\Lambda_c^{0}\,e^{- 1/(\kappa\mu)} + \Lambda_{+}^{0}\,e^{-q_{+}/(q_{c}\kappa\mu)} 
= \Lambda_{-}^{0}\,e^{+q_{-}/(q_{c}\kappa\mu)}.
\end{equation}
This, together with the normalization, gives 
$\Lambda_{+}^{0} = \frac{1}{2}\frac{q_{c}^2}{q_{+}^2}(\kappa\mu)^2$ 
and $\Lambda_{-}^{0} = e^{-(q_{+} + q_{-})/q_{c}\kappa\mu}\Lambda_{+}^{0}$, showing 
that negative ion density vanishes exponentially fast as $\kappa\mu \rightarrow 0$. 
It is not surprising  that the positive salt-ion density shows a similar functional 
dependence on the distance from the charged wall as the counterion density. 

Let us now consider the validity of the SC-DH theory by 
comparing the leading order counterion distribution 
eq.~(\ref{eq:density_c0normal_1plate}) with the next leading 
contribution eq.~(\ref{eq:density_c1normal_1plate}). 
This gives a validity condition in terms of 
$\tilde{z}^2 < 2\Xi/(1 - \kappa\mu\Xi)$, that makes sense 
as long as $\kappa\mu < \Xi^{-1}$. 
This means in fact that the SC-DH theory is valid for 
larger distances from the wall when compared to the zero-salt case 
\cite{netz:sc}. This is clearly in accord with the features of the phase 
diagram in fig.~\ref{fig:phase_diagram}, where by adding 
salt the SC-DH expansion eventually becomes valid for all 
separations from the wall.
In the regime $\Xi^{-1} < \kappa\mu$, this validity condition 
is always satisfied. In reality, this regime is easily applicable, 
since this condition can be written as $q_{c}^2\kappa \ell_{\mathrm{B}} > 1$. 
In physiological salt conditions with $\ell_{\mathrm{B}}\simeq 0.7{\mathrm{nm}}$ and $\kappa\simeq 1{\mathrm{nm}}^{-1}$, this reads as $0.7q_c^2 > 1$, 
which is already satisfied for divalent counterions.

Finally, in the regime $\kappa\mu > 1$ we need to generalize 
the calculation to take into account corrections of order $(\kappa\mu)^2$ 
to the next leading counterion density $\rho_{c}^{1}(\tilde{z})$ 
in eq.~(\ref{eq:density_c1normal_1plate}). This calculation will be 
presented in the forthcoming publication \cite{olli:sc}.

Furthermore, one has to notice that eq.~(\ref{eq:density_c1normal_1plate}) 
holds only in the case when the interaction between positive ions 
and negative salt ions does not give any significant contribution 
to the densities. These plus-minus attractions induce an extra length-scale 
to the problem, the radius of ions $a$, which has to be non-zero to 
cut off these interactions. In the limit $\kappa\mu \rightarrow 0$, i.e. 
when the interaction between ions is assumed to be unscreened, 
the results derive above hold as long as 
\begin{equation}
\label{eq:criterion_1_order}
\kappa\mu < \frac{2\tilde{a}}{\Xi} + \mathcal{O}(\ln[\kappa \ell_{\mathrm{B}}/(\kappa a)^2]), 
\end{equation}
where $\tilde{a}$ is the rescaled ion-radius \cite{olli:sc}. 
This clearly means, that we cannot reduce $a$ to zero without 
removing all the salt, i.e. setting also $\kappa\mu=0$. This is clearly 
caused by the fact, that even in the presence of very small amount 
of negative salt ions the Mayer functions of oppositely charged 
ions start to diverge, indicating complexation into Bjerrum pairs.

In summary, the  SC-DH limit, where typically one has $\tilde{a} = a/\mu \gg 1$, 
thus applies to the leading order for all values of 
$\tilde{z}$ , as long as $\Xi^{-1} < \kappa\mu <  2\tilde{a}/\Xi$.
The criterion eq.~(\ref{eq:criterion_1_order}) does not necessarily 
imply that the SCDH limit becomes invalid for larger values of $\kappa\mu$. 
If $\kappa\mu$ exceeds the value given by eq.~(\ref{eq:criterion_1_order}), 
one has to take into account the screening of 
interactions even between positive and negative salt ions themselves, thus
making SCDH theory applicable for all values of salt concentrations \cite{olli:sc}.

The presence of these screened interactions increases the density of positive ions close
to a charged wall and decreases the density of negative salt
ions \cite{olli:sc}. Intuitively, this is the case because single-particle density
for positive ions is a decreasing and for negative ions
an increasing function of distance from the wall. In the SC-limit, the plus-
minus interaction is the dominant contribution to the second
order virial expansion term. To maximize attraction
between negative and positive ions all the density profiles become
more sharply peaked close to the wall. However, this happens in such a way that electroneutrality
holds, which furthermore
means that the integral over the total charge density has to
vanish to all higher orders. As a consequence, close to the
wall, the amount of positive ions increases, and the amount
of negative ions decreases. Clearly, the opposite happens
far away from the wall.

The second important contribution 
to next-leading order densities arises from the subtraction
of artificial DH salt bath. This again gives positive contribution
to positive ion density and negative
contribution to negative ion density close to the wall, and is due to the fact that one
is removing perturbatively the artificial ions involved in the screening, 
thus decreasing the
contact densities of negative ions and increasing the contact density of positive ions.

The asymmetric virial expansion presented above also applies 
to the high salt limit $\kappa\mu \rightarrow \infty$ as long as $\Xi \gg 1$. In this 
limit,  the Mayer functions become extremely 
short-ranged due to an almost complete screening of 
electrostatic interactions. 
As a consequence, one can expand these functions in $v_\mathrm{DH}$, 
giving rise to an expansion independent of $\Xi$. This expansion 
is valid only if $v_\mathrm{DH}(2a) \ll 1$, where 
$2a$ is the ionic diameter. 
 To the leading order, however, the density expressions 
derived above apply also to high salt case \cite{olli:sc}.

The present theory can be readily employed for further,
more complex applications including charged polymers and colloids
with biologically relevant concentrations of salt.
One can also explore situations involving the joint interplay of
counterions and salt within the present theoretical framework, which
will be discussed elsewhere \cite{olli:sc}.

Another phenomenon that we have not yet accounted for thus far is overcharging
and/or charge inversion observed in simulations of charged surfaces
with multivalent counterions in salt solution \cite{Tanaka,Martin-Molina,Messina-Holm,Lenz-Holm,
Jonsson-1,Jonsson-2}. This phenomenon has also been extensively
studied in several recent theoretical works \cite{shklovskii,shklovskii-2a,shklovskii-2b,Levin,Greberg}.
It is understood that ion-ion electrostatic correlations \cite{shklovskii,Levin} as well as excluded-volume
effects due to finite ion size \cite{Greberg,Messina-Holm} play essential roles in the mechanism of 
overcharging. Physically, the absence of this effect in our formalism so far
may be traced back to the limiting one-particle structure of the leading-order
SC-DH theory. A thorough analysis
of higher-order terms is required to account for many-body and
excluded-volume effects which go beyond the scope of the present paper and 
will be presented in a forthcoming article \cite{olli:sc}. 
It can be shown that the validity of the SC-DH theory strongly depends on
the value of the ion-diameter through Mayer-functions, and can
actually break down in the limit where the interaction between
oppositely charged particles exceeds $k_\mathrm{B} T$, 
which is already indicated by eq.~(\ref{eq:criterion_1_order}). 
This limit indicates
the onset of Bjerrum-pairing, thus destroying the Debye screening
picture [22]. 
Our preliminary simulations \cite{Olli-Woon} indicate that at large
coupling parameters overcharging is indeed possible but typically amounts to a fraction 
of the bare surface charge, which in this feature agrees with other simulations
\cite{Martin-Molina,Messina-Holm,Lenz-Holm,Jonsson-1,Jonsson-2}
and appears to reflect the second-order nature of this phenomenon.

In summary, we have derived analytic expressions for 
counter- and co-ion distributions in the presence of salt 
in the vicinity of a charged macroion. The results are 
consistent with previous work in the PB, DH and 
SC limits and fill the gap between the no-salt and high salt 
limits. The results presented here are relevant in a multitude 
of soft-matter and biological systems characterized by 
non-negligible salt concentrations, and pave the way for 
further applications \cite{olli:sc}, complementing those 
presented here.

{\em Acknowledgement -- } 
P.\ L.\ H.\ would like to thank the Danish 
National Research Foundation for its financial support 
in the form of a long-term operating grant awarded to 
MEMPHYS--Center for Biomembrane Physics. 
This work has, in part, been supported by the Academy of 
Finland, Finnish Cultural Foundation, and the Danish National
Research Foundation.


\end{document}